\documentclass[a4paper,aps,pre,twocolumn,groupedaddress,showkeys,showpacs,floats,floatats,floatfix]{revtex4-1} 
\usepackage[utf8]{inputenc}
\usepackage[english]{babel}
\usepackage{graphicx}
\usepackage{dcolumn} 
\usepackage{bm}
\usepackage{natbib}
\usepackage{latexsym}
\usepackage{mathrsfs}
\usepackage{amssymb}
\usepackage{amsmath}
\usepackage{amscd}
\usepackage{color}
\usepackage{pifont}
\usepackage{pstricks,pst-node,pst-text,pst-3d}
\usepackage{verbatim}
\usepackage[T1]{fontenc}

\begin{document}  

\title{Comment on "Inflation with a graceful exit and entrance driven by Hawking radiation"}

 \author{Javad T. Firouzjaee}
\affiliation{ School of Astronomy and School of Physics, Institute for Research in Fundamental Sciences (IPM), Tehran, Iran }
 \email{j.taghizadeh.f@ipm.ir}

%\date{\today}

\begin{abstract}
Modak and Singleton [Phys. Rev. D 86, 123515 (2012)] have presented Hawking-like radiation for cosmological inflation which has a natural ``turn on" and a natural ``turn off" mechanism. This Hawking-like radiation results in an effective negative pressure ``fluid" which leads to a rapid period of expansion in the very early Universe. We discuss that the turn on mechanism can not happen for FRW model in early universe because its horizon is apparent horizon not event horizon. Hence, we cannot apply geometric optic approximation which is a necessary condition for tunneling method. It was shown that this model predict a value for $\frac{\rho}{m_{pl}^4}$  which is bigger than the COBE normalization constraint in the Cosmic Microwave Background (CMB) at the horizon exit.

\end{abstract}

\pacs{98.80.Cq}

\maketitle

In the recent paper \cite{modak} the Authors claim that they propose a mechanism for inflation based on the particle creation due to Hawing radiation in a FRW space-time. Core of their calculation is based on the associating the temperature for the FRW apparent horizon \cite{cai}.
This method is called Hamilton-Jacobi or tunneling method \cite{tunneling}. 
The Hamilton-Jacobi method to calculate the Hawking radiation uses the fact that
within the WKB approximation the tunneling probability for the
classically forbidden trajectory from inside to outside the
horizon is given by
\begin{equation}
\Gamma \propto \exp\left(- \frac{2}{\hbar}\mbox{Im } S\right),
\label{prob}
\end{equation}
where $S$ is the classical action of the (massless) particle to the
leading order in $\hbar$ \cite{tunneling}.
It was shown \cite{javad} that only in the case of the de Sitter space which FRW apparent horizon is event horizon, we can write the WKB or geometric optics approximation for the horizon and apply the tunneling method.
There is a confusing point that one can attribute a temperature to the FRW apparent horizon and write the area law for it \cite{cai}, but this does not mean that this system has the Hawking radiation.\\

Furthermore, if we use the canonical invariance tunneling method \cite{zhu}, we are not allowed to write the standard ansatz for scalar wave function $\phi = exp[-\frac{i}{\hbar} S(r,t)+...]$ and taking the limit as $\hbar \sim 0 $ for the FRW apparent horizon in the radiation dominated era, and getting the Hamilton-Jacobi equation. Because this ansatz can be written in the case that we are near the (event) horizon \cite{padmanabhan}. In the case near (event) horizon, the WKB or geometric optics approximation for the wave is satisfied and we can write this ansatz for the scalar wave function.\\

Even in the deriving Hawking radiation from the Bogolubov  coefficient \cite{radiation}, we only able to drive radiation from slowly varying space time or the space times which have adiabatic vacuum, and no one are able to define particle in non static space time such as radiation dominated FRW universe.

According to the \cite{modak}, The first law of thermodynamics can be rewritten as 
\begin{equation}
\frac{d(\rho V)}{dt}+p \frac{dV}{dt}=+\frac{dQ}{dt}=\sigma A_H T^4, \label{power}
\end{equation} 
this equation becomes
\begin{equation}
\dot{\rho} +3 (\rho + p )\frac{\dot a}{a} = \frac{3\sigma}{c}\left(\frac{\hbar}{2\pi k_{B}}\right)^4 H^5, \label{neq}
\end{equation}
after simplification, we gets
\begin{equation}
\frac{\dot{\rho}}{\rho}+3(1+\omega - \omega_c (t))\frac{\dot a}{a} = 0, \label{neq1}
\end{equation} 

where $p = \omega \rho$ and the
particle creation equation of state is 
\begin{equation}
p_c(t)=\omega_c(t) \rho ~.
\label{crph}
\end{equation}
We can write the equation of the state in the form
\begin{equation}
\label{omega-c} 
\omega_c (t) = \alpha \rho (t) ~~, {\rm where}  ~~\alpha = \frac{\hbar G^2}{45 c^7}= 4.8 \times 10^{-116} (J/m^3)^{-1}
\end{equation} 
This negative pressure that occurs due to Hawking radiation in FRW space-time could give the inflationary era.  More precisely, when the energy density of the universe is near Planck energy density,  $10^{-114} (J/m^3)$, the Hawking radiation term become important and act as a negative pressure which cause an accelerated universe.  As the universe expand, this term becomes negligible and we will have a smooth transition to radiation dominated universe.  \\
Since we can not write the WKB approximation for the FRW universe, we don't have any negative pressure that create the expanding universe and therefore no inflationary phase.\\

\textbf{Can this model  be consistent with CMB observation?}

Having inflationary mechanism from Hawking like radiation method lead to
\begin{equation}
w_c(t)= \frac{\hbar G^2}{45 c^7} \rho \simeq 4/3.
\label{ffff}
\end{equation}
or  in the God-given natural units $\dfrac{\rho}{m_{pl}^4} \simeq 1$. This require that this mechanism of inflation be  near-Planck-scale physics rather than grand unified scale. \\

On the other hand, cosmological perturbation theory says that
\begin{equation}
P_t = r  P_R 
\end{equation}
in which $r$ is the ratio of the tensor mode to the scalar mode.
The upper bound on tensor perturbations from WMAP and PLANCK 
implies that  r<0.11.  Now, from the COBE normalization for the curvature 
perturbations at the horizon exit point we have $ P_R \simeq 10^{-9}$. As a result   
$P_t \simeq \frac{H^2}{m_{pl}^2} \simeq \frac{\rho}{m_{pl}^4}  < 10^{-10}$ \cite{bassett}. If we had  (\ref{ffff}), then the tensor perturbation must be seen in the Cosmic Microwave Background. Therefore, this model model predict a value for  $\dfrac{\rho}{m_{pl}^4} \simeq 1$ which is bigger than the predicted value for the COBE normalization for the curvature perturbations at the horizon exit point .\\

\textbf{Acknowledgment}
I would like to thank Hassan Firouzjahi and Mohammad Hossein Namjoo for useful discussions and comments. I wish to thank the Douglas Singleton for his useful answers for my questions.

\end{document}